\documentclass[preprint,amsmath,amssymb]{revtex4}

\usepackage{graphicx}
\usepackage{dcolumn}
\usepackage{bm}
\usepackage{multirow}
\usepackage{amsmath}
\usepackage[version=3]{mhchem}
\usepackage{float}

\def\mathbi#1{\textbf{\em #1}}

\usepackage[dvips]{color}
\definecolor{darkgreen}{rgb}{0,.6,0}

\begin{document}
\title{Role of Water Molecule in Enhancing the Proton Conductivity on Graphene Oxide at Humidity Condition}

\author{Gum-Chol Ri, Jin-Song Kim, and Chol-Jun Yu\footnote{E-mail: ryongnam14@yahoo.com}}
\affiliation{Department of Computational Materials Design, Faculty of Materials Science, Kim Il Sung University, Ryongnam-Dong, Taesong District, Pyongyang, Democratic People's Republic of Korea}
\date{\today}

\begin{abstract}
Recent experimental reports on in-plane proton conduction in reduced graphene oxide (rGO) films open a new way for the design of proton exchange membrane essential in fuel cells and chemical filters. At humidity condition, water molecules attached on the rGO sheet are expected to play a critical role, but theoretical works for such phenomena have been scarcely found in the literature. In this study, we investigate the proton migration on water-adsorbed monolayer and bilayer rGO sheets using first-principles calculations in order to reveal the mechanism. We devise a series of models for the water-adsorbed rGO films as systematically varying the reduction degree and water content, and optimize their atomic structures in reasonable agreement with the experiment, using a density functional that accounts for van der Waals correction. Upon suggesting two different transport mechanisms, epoxy-mediated and water-mediated hoppings, we determine the kinetic activation barriers for these in-plane proton transports on the rGO sheets. Our calculations indicate that the water-mediated transport is more likely to occur due to its much lower activation energy than the epoxy-mediated one and reveal new prospects for developing efficient solid proton conductors.
\end{abstract}
\maketitle

\section{Introduction}
Proton exchange membrane is an essential component in electrochemical energy generating and storage devices such as fuel cells and batteries as well as selective material sieving systems like sensors and chemical filters. To date, Nafion and Nafion-based artificial materials have been most widely used as an efficient proton exchange membrane, but they have serious problems of high cost and conductivity loss at the temperature over 80 $^\circ$C~\cite{Mauritz,Lee,Nicotera}. Recently, graphene oxide (GO) and reduced GO (rGO) films with a controlled reduction degree have attracted considerable attention as a superior solid electrolyte for proton exchange to Nafion due to their low cost, easy fabrication and environmental friendliness~\cite{Joshi,Kim,Li,Holmes}.

It was known that proton can pass across monolayer graphene~\cite{Hu,Walker,Goh} or few layer GO sheet~\cite{Joshi,Kim,Li}, but only through atomic-scale defects created on those nanosheets. If no defect, graphene and GO sheet are factually impermeable to proton under ambient condition due to a dense, delocalized electronic cloud formed by the $\pi$-orbitals of graphene~\cite{Tsetseris,Berry}. In fact, Hu {\it et al}.~\cite{Hu} measured an areal conductivity of proton across the monolayer graphene sheet as $\sigma=S/A\approx$ 2$\sim$4 mS/cm$^2$ ($S=I/V$ is the conductance and $A$ is the area of the sample) at room temperature with the corresponding activation barrier of $E_a\approx0.78$ eV. For such proton transport, first-principles calculations yielded even higher values of 1.25$\sim$1.56 eV due to different proton transport pathways from real experiment such as transport in vacuum rather than in aqueous environment~\cite{Miao,WangKax,Kroes,Shi}. Decorating graphene with catalytic metal nanoparticles like Pt slightly reduces the activation barrier as much as $\sim$ 0.5 eV~\cite{Hu}, which is still relatively high. Therefore, research effort has been focused on creating nanopores with precisely controlled narrow size distributions on multilayer graphene or GO films to achieve easy pass of proton~\cite{TYang}. In these porous multilayer films, protons can move in one layer and pass to another through nanopores~\cite{Hatakeyama3}. Then, the problem changes from the through-plane conductivity to the in-plane one, but the activation barrier for proton or hydrogen atom transport on the graphene sheet was turned out to be high as well, like 0.9 eV due to a strong binding of H to graphene sheet~\cite{Zhao}.

Unlike graphene, GO has oxygenated functional groups such as epoxy ($-\text{O}-$) and hydroxy ($-\text{OH}$) groups, which form one-dimensional hydrogen-bonded channels for proton transport~\cite{Sljivancanin,Wang,Topsakal,Dimiev,Raidongia}. Recently, Hatakeyama {\it et al.} reported that multilayer GO films have a good in-plane proton conductivity at room temperature and high relative humidity (RH) condition~\cite{Hatakeyama1,Hatakeyama14,Hatakeyama15,Karim,Wakata,Koinuma}. They measured the in-plane proton conductivity of rGO film to be $\sigma=(S\times L)/(T\times D)\approx$ 2.4 mS/cm ($L$, $T$, and $D$ are the width, thickness, and length of the sample) at 278 K and 90\% RH with the corresponding activation barrier of 0.12 eV. When added some functional groups to GO, forming {\it e.g.} GO-Nafion hybrid~\cite{Tseng,Nicotera} or sulfonated GO complexes~\cite{Hatakeyama1,Hatakeyama15,Wakata,Zhaosul}, the in-plane proton conductivity was observed to be enhanced. These experimental findings imply that uptake and retention of water in the GO sheet are a key factor for the high proton conductivity; even for the case of graphene the activation barrier reduces by 0.42 eV when mediated by water molecules~\cite{Zhao}. It was expected that the hydrophilic functional groups such as epoxy groups in the GO sheet can readily adsorb water molecules and support the channel formation for proton stream on the sheet. In this context, it is urgent to theoretically reveal the mechanism behind enhancing proton transport by water adhesion to GO sheet for the design of novel functional GO-based solid electrolyte. To the best of our knowledge, however, theoretical works for these phenomena have been scarcely reported, although there exist first-principles works for proton penetration through graphene and other 2D materials~\cite{Hu,Miao,WangKax,Kroes,Shi}.

In this work, we investigate the atomic structures of water-adsorbed monolayer and bilayer rGO sheets and proton migrations on these sheets by using first-principles method within density functional theory (DFT) framework. The van der Waals (vdW) dispersive interactions between the graphene sheets and molecules are included using the flavor of vdW-DF-OB86~\cite{vdWDFob86}. We predict the migration paths of proton on the rGO sheets by estimating the bond valence sum (BVS)~\cite{Brown_BVS}, and calculate the activation barriers for these in-plane proton migrations by using the climbing image nudged-elastic-band (NEB) method~\cite{NEB}. Based on the calculation data, we propose the most reasonable mechanism behind the enhancement of proton conductivity by water molecules.

\section{Computational Methods}
We make the atomistic modelling of rGO sheets and build the corresponding supercells. For the types of rGO with different oxidation degrees ({\it i.e.}, reduction degree), we consider both monolayer and bilayer rGO sheets with gradually increasing O/(C+O) ratios from the minimum value of 14.3\% to the maximum value of 33.3\% in this study. The orthogonal ($7\times3$) and ($4\times2$) cells are employed for the basal plane of graphene sheet, which contain the carbon atoms of 72 and 32, respectively. On both sides of graphene sheet, the same number of oxygen atoms are adsorbed to form epoxy groups. Here, we arrange the epoxy groups as a continued row to form a migration path for the in-plane proton transport, based on the established fact that the epoxy groups are clustered on the rGO sheet~\cite{Sljivancanin}. Then, the numbers of oxygen atoms can be 12 for the case of ($7\times3$) cell and 8, 12, 16 for the case of ($4\times2$) cell, which correspond to the chemical formula of \ce{C72O12} (O/(C+O) = 14.3\%), \ce{C32O8} (20.0\%), \ce{C32O12} (27.3\%) and \ce{C32O16} (33.3\%). The simulated lattice constant of 2.46 \AA~and the vacuum layer of 15 \AA~thickness are used throughout the work.

All calculations in this work are performed using the pseudopotential plane-wave method as implemented in Quantum ESPRESSO package (version 5.3)~\cite{QE}. We use the Vanderbilt-type ultrasoft pseudopotentials to describe the interaction between ions and valence electrons~\footnote{We used C.pbe-van\_ak.UPF, O.pbe-van\_ak.UPF, and H.pbe-van\_ak.UPF, which are provided in the package.}. The Perdew-Burke-Ernzerhof (PBE) functional~\cite{PBE} within the generalized gradient approximation (GGA) is used for the exchange-correlation interaction between valence electrons. For the vdW dispersive interaction between the graphene sheets and water molecules, the vdW energy provided by vdW-DF-OB86 method~\cite{vdWDFob86} is added to the DFT total energy. As the major computational parameters, the plane-wave cutoff energies are set to be 40 Ry for wave function and 400 Ry for electron density, and the Monkhorst-Pack special $k$-points are set to be ($2\times2\times5$) for bilayer sheet and ($2\times2\times1$) for monolayer sheet, providing a total energy accuracy of 5 meV per carbon atom. Self-consistent convergence threshold for total energy is $10^{-9}$ Ry, and the convergence threshold for atomic force in structural relaxations is $8\times10^{-4}$ Ry/Bohr. Methfessel-Paxton first-order spreading with the gaussian spreading factor of 0.2 Ry is applied to the Brillouin-zone integration.

\begin{figure}[!b]
\begin{center}
\includegraphics[clip=true,scale=0.5]{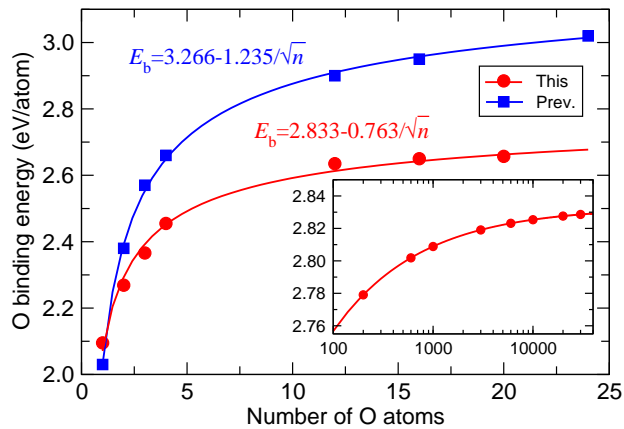}
\end{center}
\caption{\label{fig_bind}Oxygen binding energy as increasing the number of oxygen atoms in reduced graphene oxide. Inset shows an extension to over 100 oxygen atoms using the interpolation curve. Blue line shows the previous theoretical result~\cite{Sljivancanin}.}
\end{figure}

We calculate the oxygen binding energy as a function of the number of oxygen atoms in rGO and compare with the previous data from first-principles calculation~\cite{Sljivancanin} in order to check the validity of the computational parameters and our rGO supercell models. The oxygen binding energy per atom $E_b$ can be calculated as follows,
\begin{equation}
E_b=-\frac{1}{N_\text{O}}(E_\text{rGO}-E_\text{G}-N_\text{O}E_\text{O})
\end{equation}
where $E_\text{rGO}$, $E_\text{G}$ and $E_\text{O}$ are the total energies of rGO, graphene and isolated oxygen atom, and $N_\text{O}$ is the number of oxygen atoms involved in the rGO model. The result is shown in Fig.~\ref{fig_bind}. Similarly, water binding energy per molecule can be calculated as follows,
\begin{equation}
E_b=-\frac{1}{N_{\ce{H2O}}}(E_\text{hyd-rGO}-E_\text{rGO}-N_{\ce{H2O}}E_{\ce{H2O}})
\end{equation}
where $E_\text{hyd-rGO}$ and $E_{\ce{H2O}}$ are the total energies of water-adsorbed rGO supercell and \ce{H2O} molecule, and $N_{\ce{H2O}}$ is the number of water molecules. The proton adsorption energy into the water-adsorbed rGO sheet can be calculated as follows,
\begin{equation}
E_\text{ad}=E_\text{p-hyd-rGO}-E_\text{hyd-rGO}-\frac{1}{2}E_{\ce{H2}}
\end{equation}
where $E_\text{p-hyd-rGO}$ and $E_{\ce{H2}}$ are the total energies of proton-adsorbed hydrous rGO and \ce{H2} molecule.

\begin{figure*}[!t]
\scriptsize
\begin{center}
(a)\includegraphics[clip=true,scale=0.4]{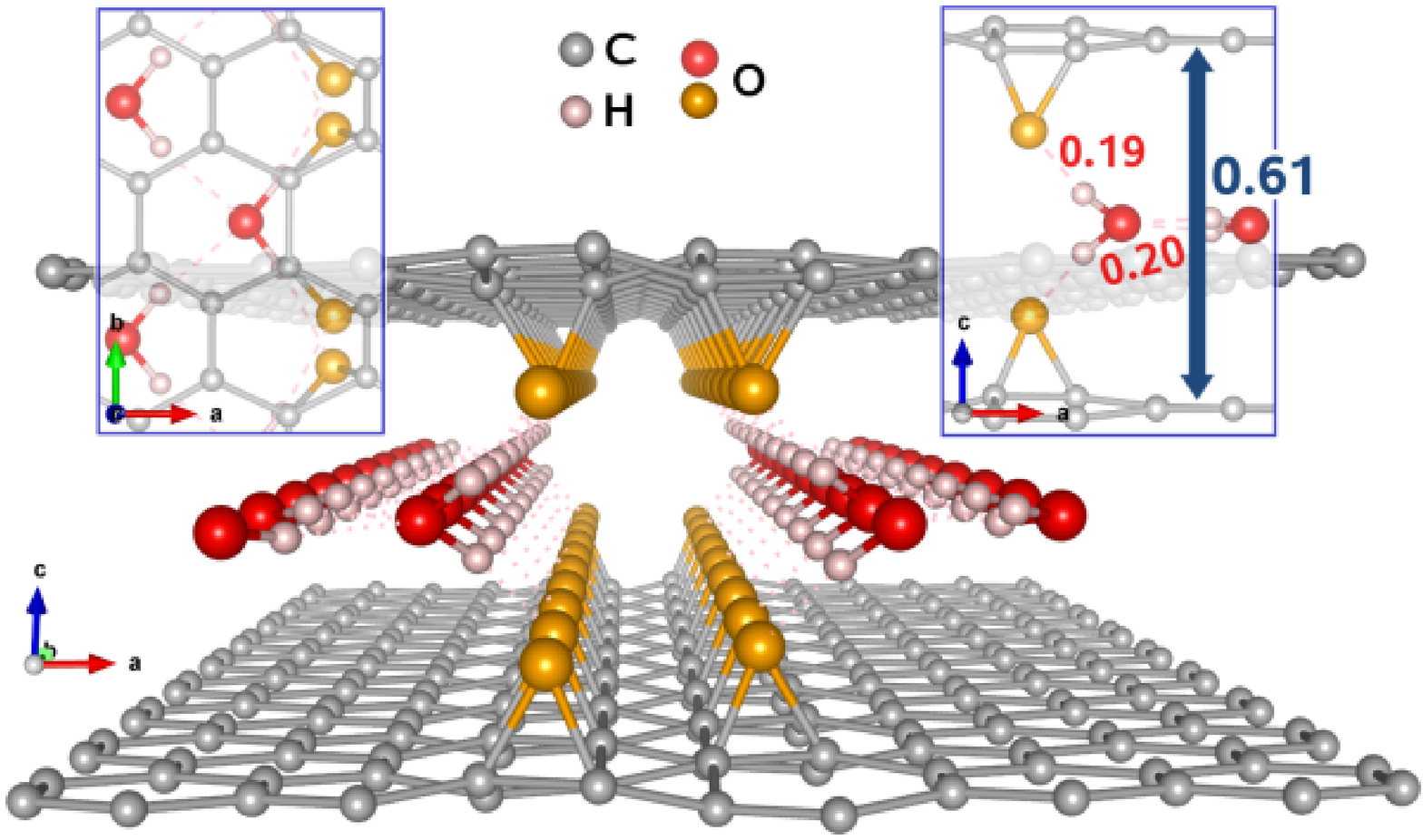}
(e)\includegraphics[clip=true,scale=0.47]{fig2e.eps} \\ \vspace{10pt}
(b)\includegraphics[clip=true,scale=0.33]{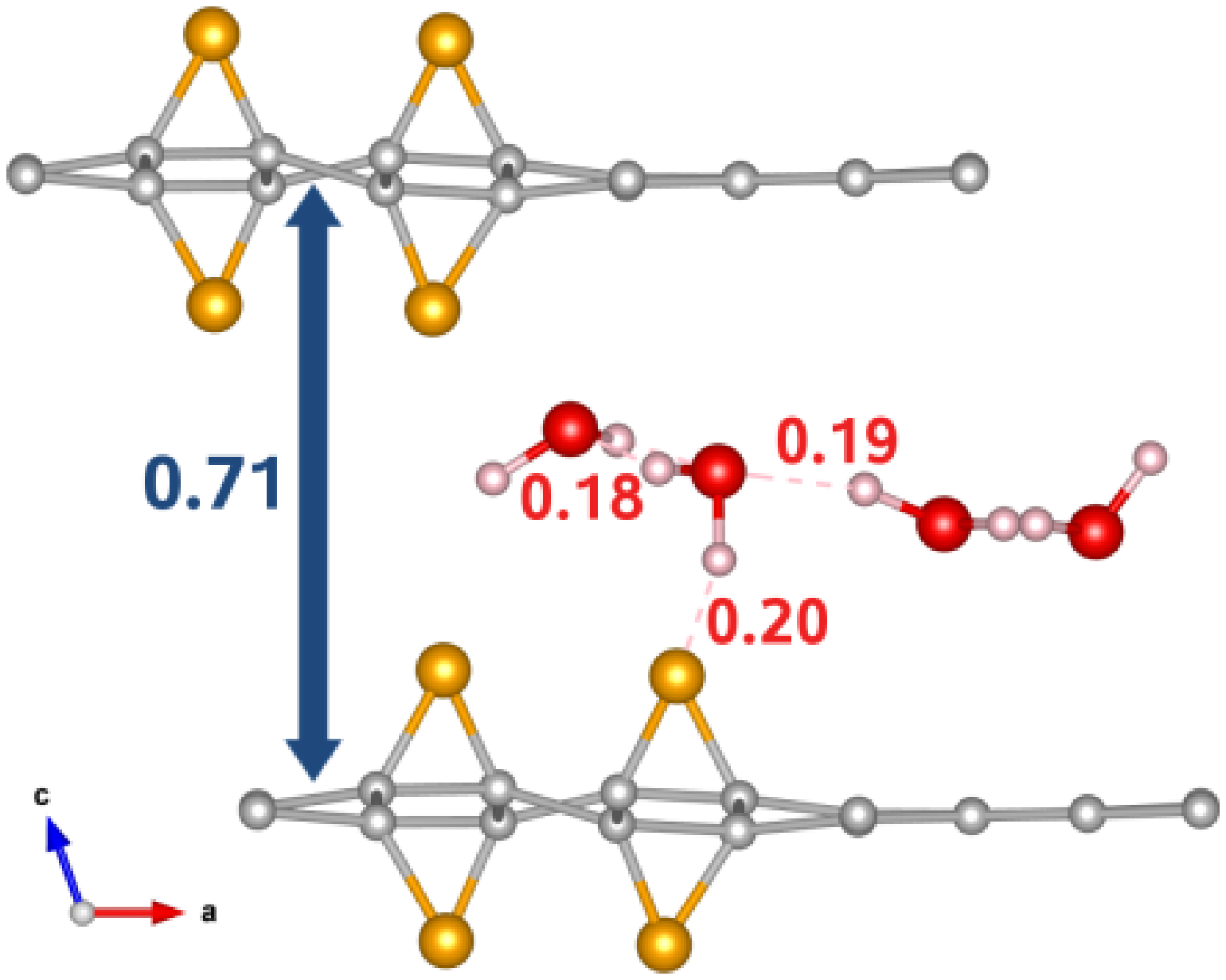}
(c)\includegraphics[clip=true,scale=0.33]{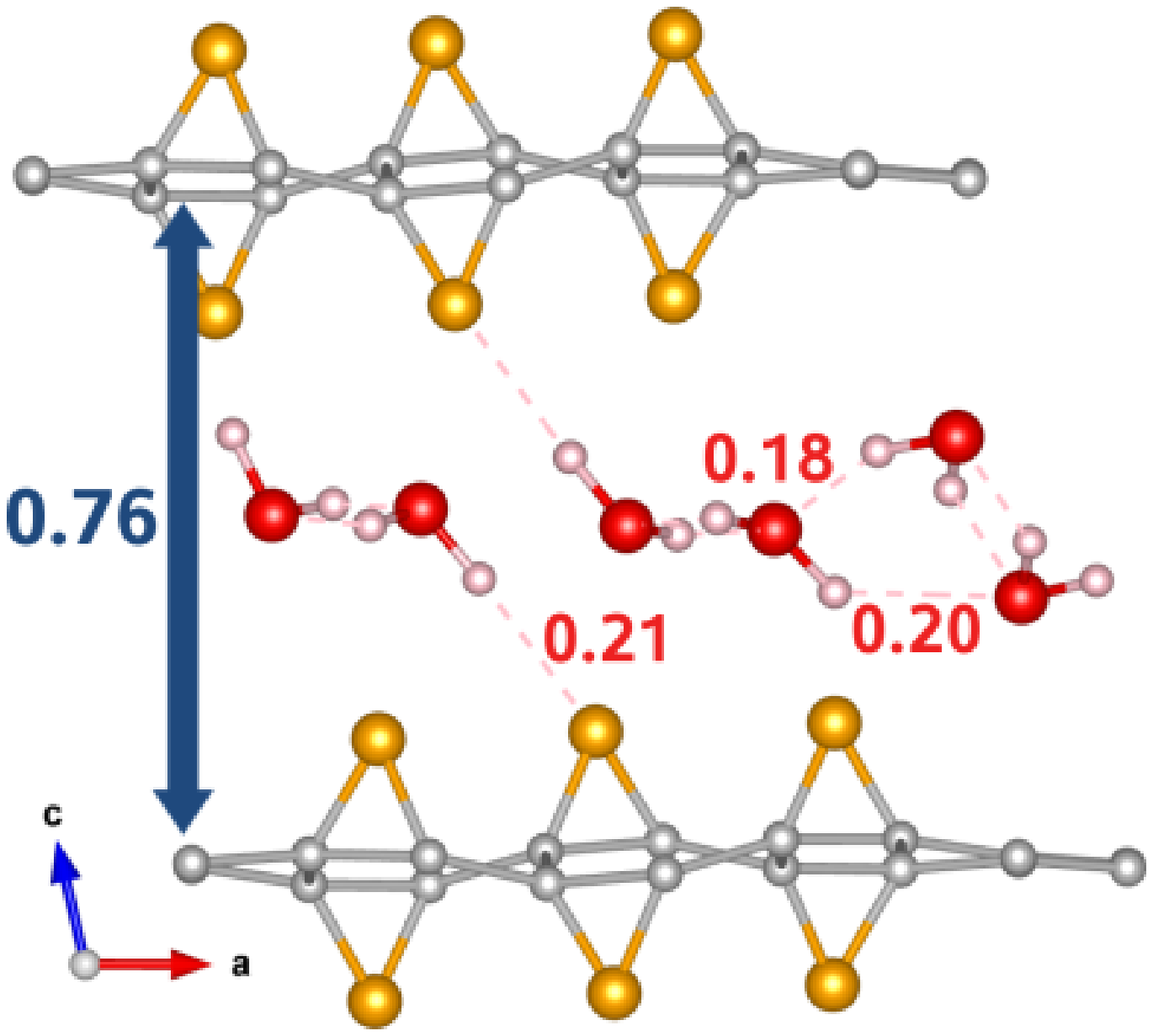}
(d)\includegraphics[clip=true,scale=0.33]{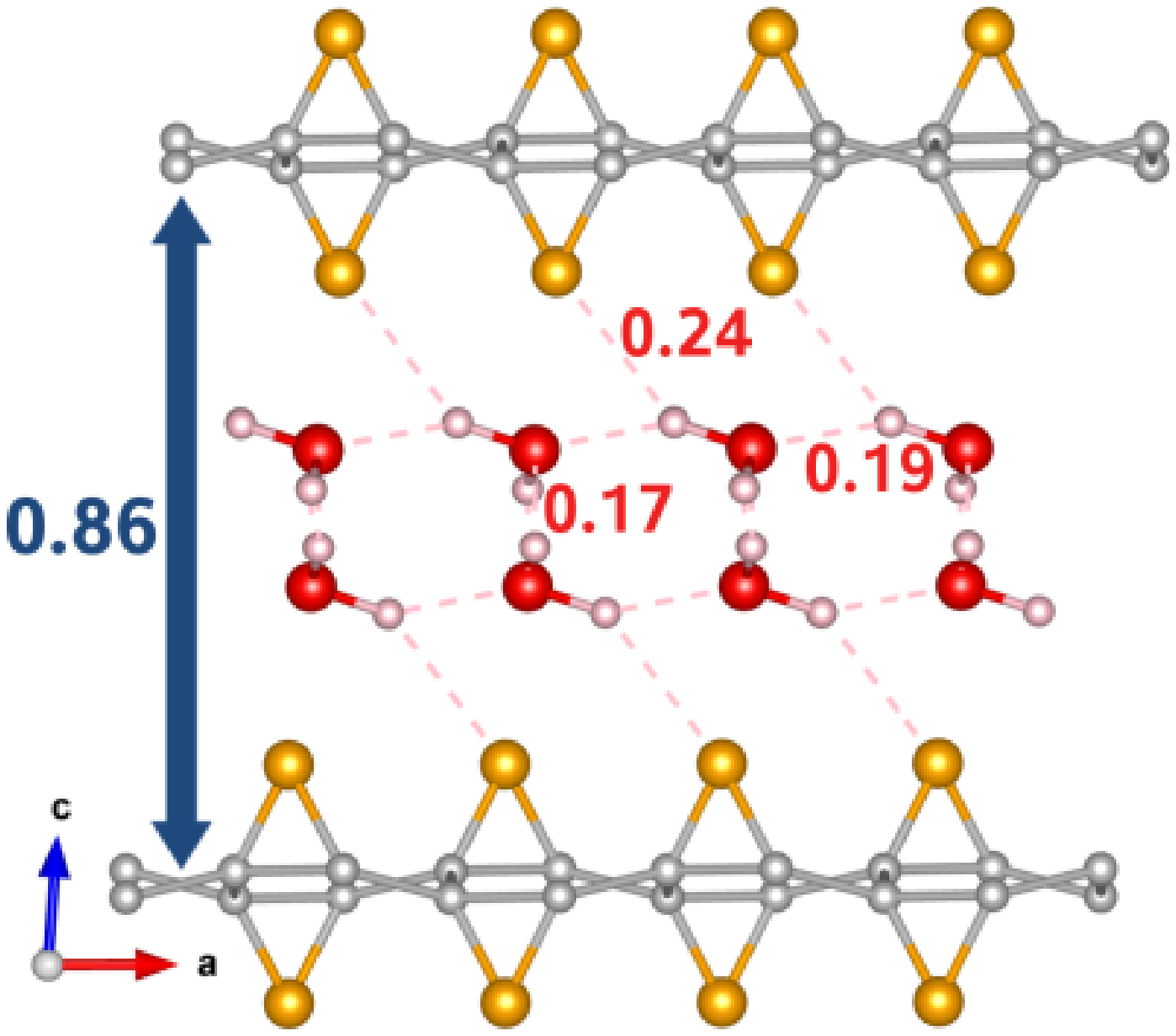}
\end{center}
\caption{\label{fig_mostr}Optimized atomic structures of water-intercalated bilayer rGO with chemical formula of (a) \ce{C72O12\cdot}12\ce{(H2O)} in perspective (center), top (left) and side (right) views, (b) \ce{C32O8\cdot}8\ce{(H2O)}, (c) \ce{C32O12\cdot}12\ce{(H2O)}, and (d) \ce{C32O16\cdot}16\ce{(H2O)} in side view. Interlayer distance and hydrogen bond length are shown in the unit of nm. (e) Interlayer distance $d$ as a function of oxidation degree (O/(C+O)(\%)), where Exp. means the experimental values in Ref.~\cite{Hatakeyama14} and Cal. the calculated values in this work. Inset shows the interlayer distance and water binding energy per molecule ($E_b$) as functions of water content (wt.\%).}
\end{figure*}

To predict possible positions of proton inserted into the water-adsorbed rGO sheets, we apply the BVS method~\cite{Brown_BVS}. The BVS at position $\mathbi{r}$, $B(\mathbi{r})$, can be calculated as follows,
\begin{equation}
B(\mathbi{r})=\sum_i \exp\left[\frac{R_0-R_i(\mathbi{r})}{b}\right]
\end{equation}
where $R_i(\mathbi{r})=|\mathbi{r}-\mathbi{R}_i|$ ($\mathbi{R}_i$ for oxygen positions),  $b$ (= 0.37 \AA) is a constant, and $R_0$ is a constant specific to the pair of hydrogen and oxygen atoms. The constant $R_0$ is estimated using the parameters of atomic size ($r$) and electro-negativity ($c$) as follows~\cite{Keeffe},
\begin{equation}
R_0=r_\text{H}+r_\text{O}-\frac{r_\text{H}r_\text{O}\left(\sqrt{c_\text{H}}-\sqrt{c_\text{O}}\right)^2}{c_\text{H}r_\text{H}+c_\text{O}r_\text{O}} 
\end{equation}
where $r_\text{H}=0.38$ \AA, $r_\text{O}=0.63$ \AA, $c_\text{H}=0.89$, and $c_\text{O}=3.15$~\cite{Bond}. The values of $B(\mathbi{r})$ at the positions of hydrogen atoms are evaluated to be almost 3 and the difference of $B(\mathbi{r})$ from this value are calculated for the whole space with a grid resolution of 0.1 \AA.

To calculate the migration barriers for the in-plane proton transport, we use the NEB method~\cite{NEB}. The supercell dimensions are fixed at the optimized supercell size during the NEB runs, while all the atoms are relaxed until the forces converge within 0.05 eV/\AA. The number of images in this work is tuned so that the distance between neighbouring NEB images is less than 1 \AA.

\begin{figure*}[!t]
\scriptsize
\begin{center}
(a)\includegraphics[clip=true,scale=0.3]{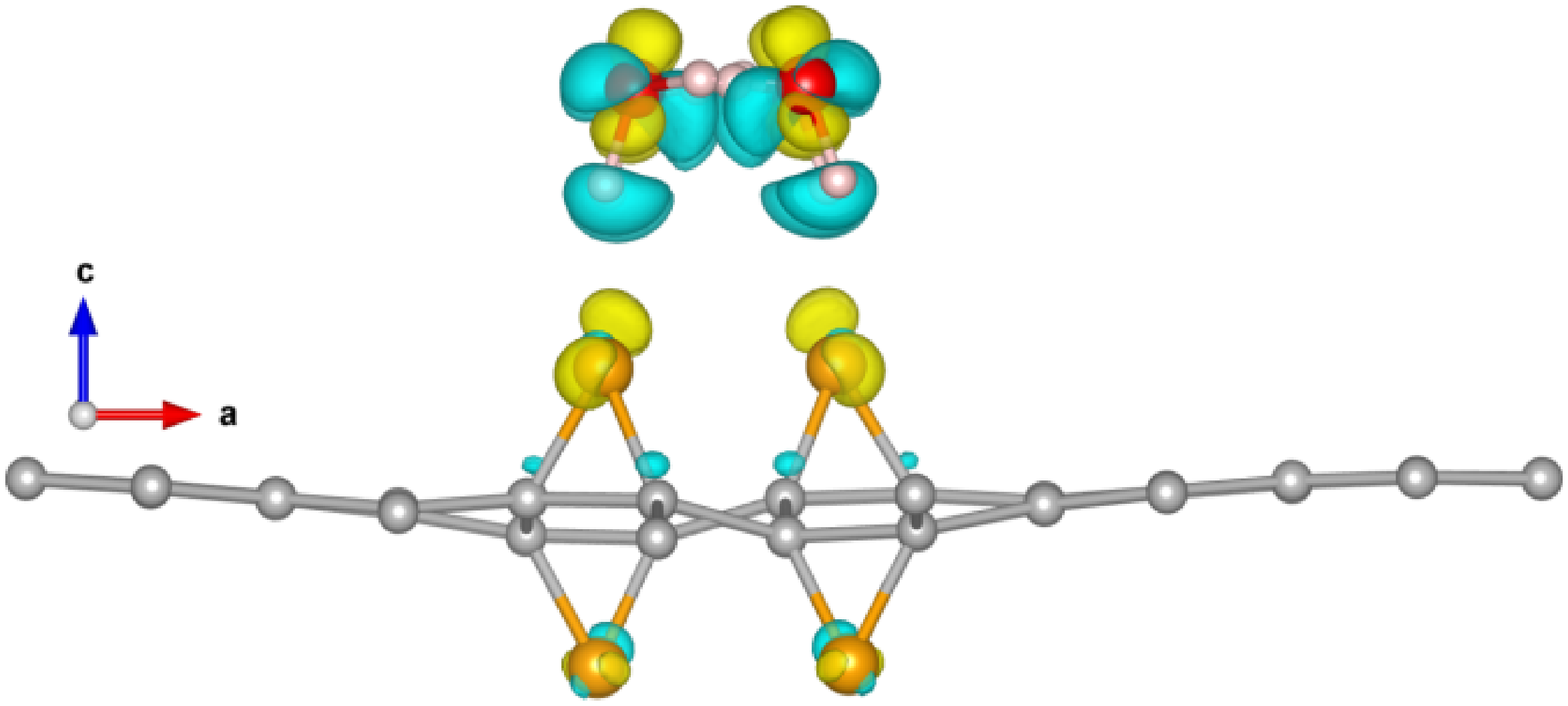}
(b)\includegraphics[clip=true,scale=0.3]{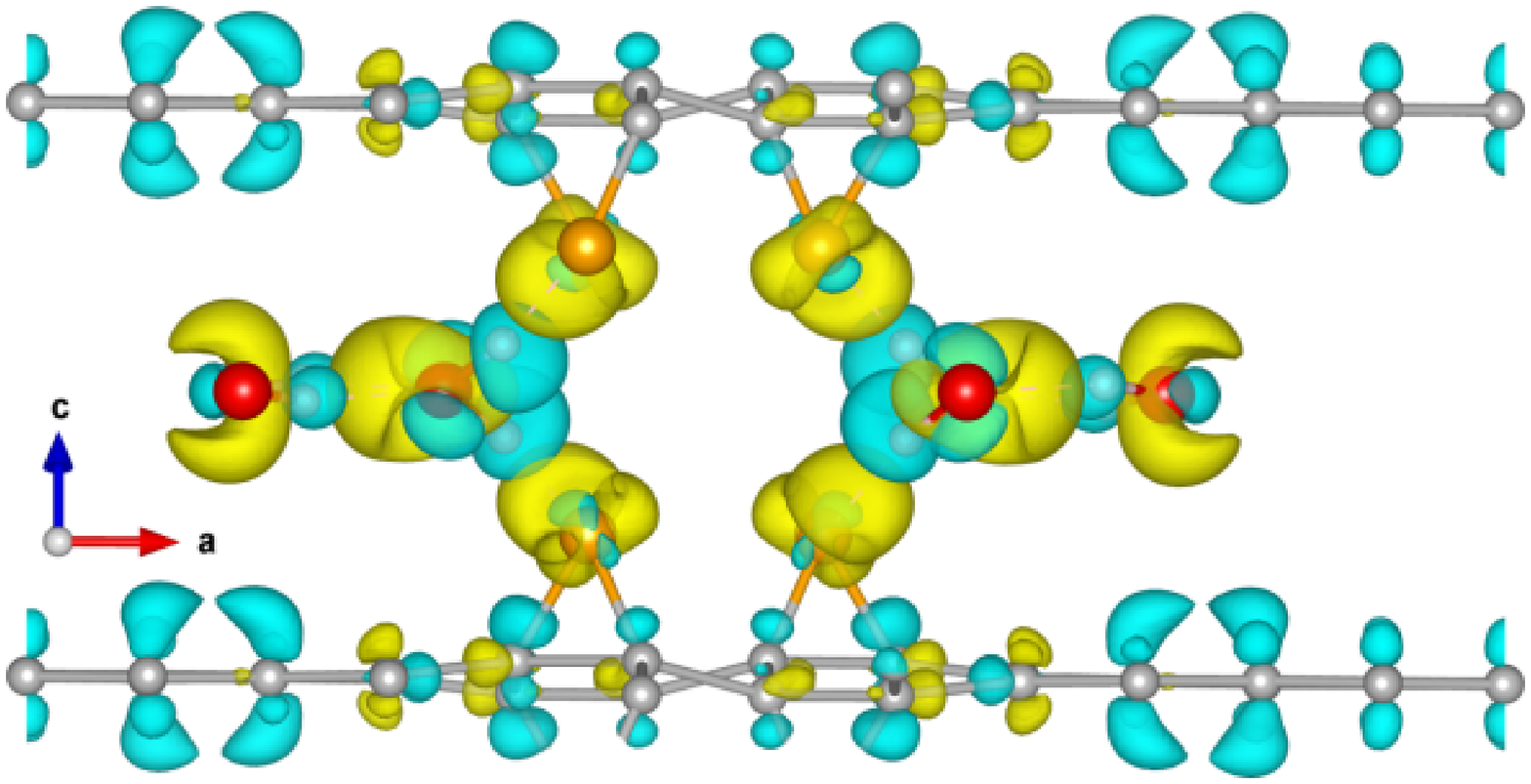} \\ \vspace{10pt}
(c)\includegraphics[clip=true,scale=0.38]{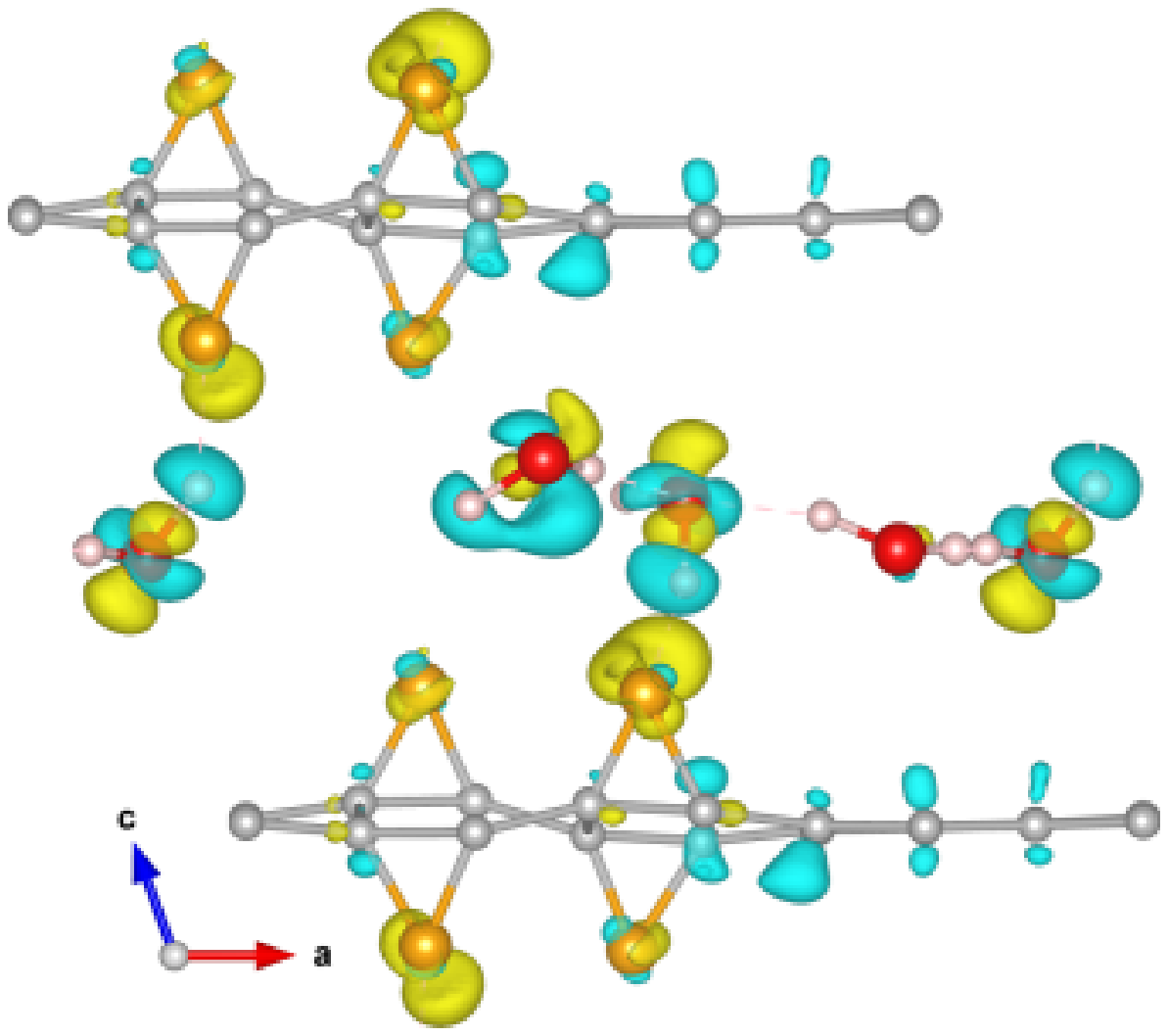}
(d)\includegraphics[clip=true,scale=0.38]{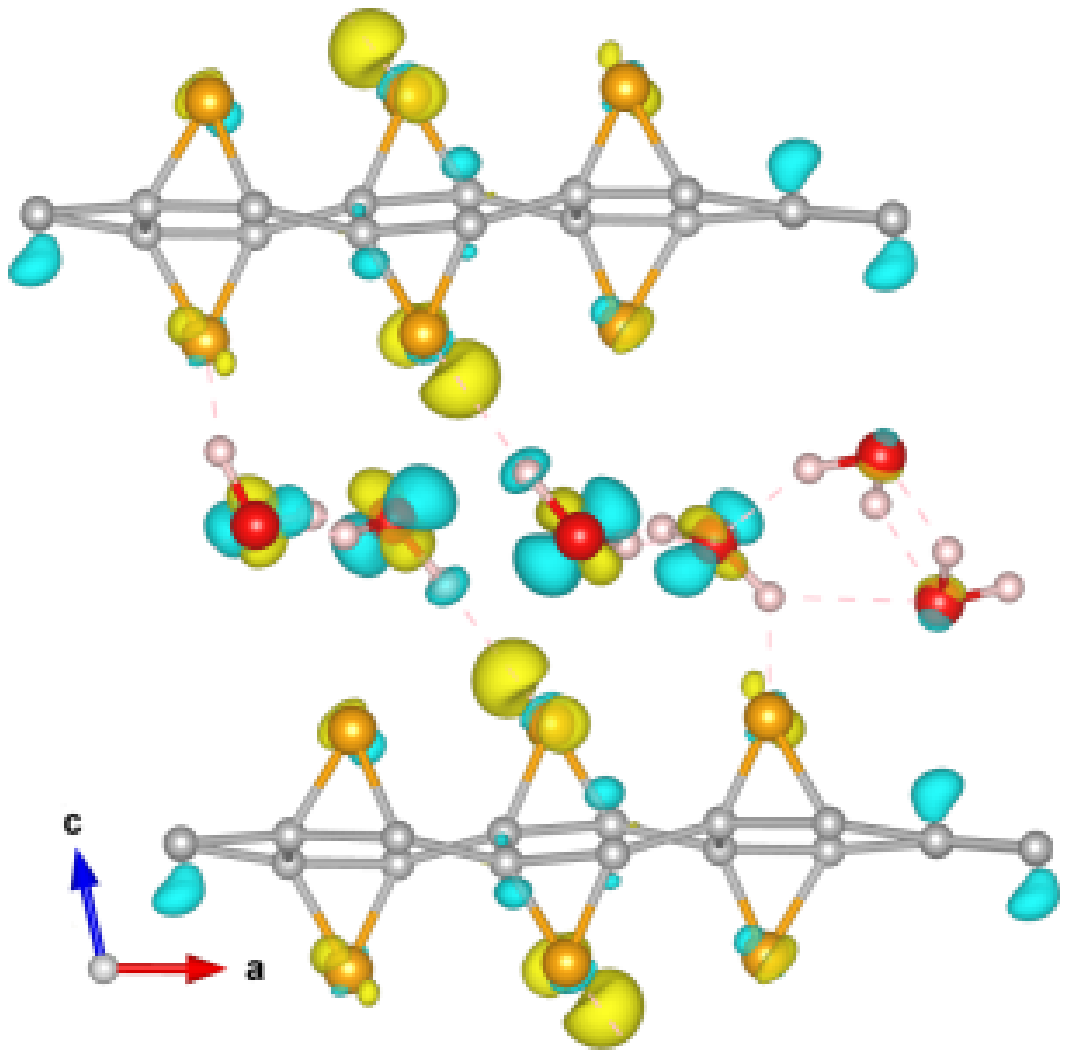} (e)\includegraphics[clip=true,scale=0.38]{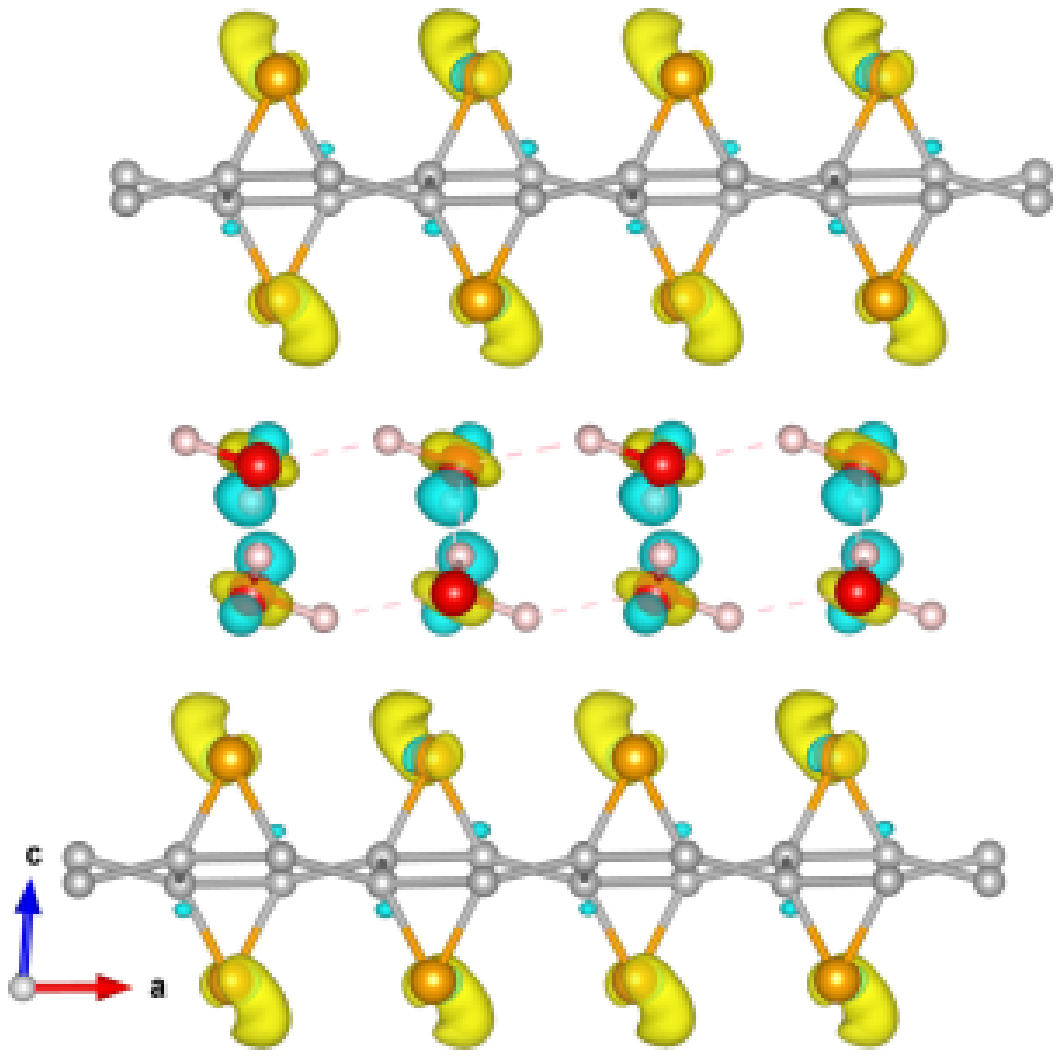}
\end{center}
\caption{\label{fig_dens}Isosurface plot of electronic charge density difference between water-adsorbed or water-intercalated rGO and pristine rGO at the value of 0.002 $|e|/$\AA$^3$. Yellow (cyan) color represents the charge accumulation (depletion). Small balls of grey, pink, brown and red colors represent the carbon, hydrogen, oxygen atoms of epoxy group and water molecule, respectively.}
\end{figure*}

\section{Results and Discussion}
\subsection{Atomic structures}
We check the validity of computational parameters and supercell models by estimating the oxygen binding energy per atom in monolayer rGO sheets with ($7\times3$) cells as increasing the number of oxygen atoms ($n=$ 1, 2, 3, 4, 12, 16, 20). It should be noted that for the cases of $n=$ 12, 16, and 20, the epoxy groups are arranged to form a low as mentioned above. We perform the atomic relaxations of these monolayer rGO supercells, and calculate the oxygen binding energies, confirming that they are agreed well with the previous results~\cite{Sljivancanin} as shown in Fig.~\ref{fig_bind}. In the special case of \ce{C72O12}, the value of 2.64 eV in this work is in reasonable agreement with the previous value of 2.90 eV. Furthermore, interpolating the calculation data into a square root function of oxygen number gives the function of $E_\text{b}(n)=2.83-0.76/\sqrt{n}$ (eV), which is comparable with the previous result $E_\text{b}(n)=3.27-1.24/\sqrt{n}$. We should emphasize that, although our calculation data are slightly underestimated compared with the previous data, possibly due to the difference of computational method and moreover inclusion of vdW correction in this work, the increasing tendencies are coincident with each other as the square root functions of oxygen number.

Then, water molecules are enforced to be adsorbed on the monolayer rGO sheet or intercalated into the interlayer space in the bilayer rGO sheets. The intercalated water molecules are placed on the center of carbon hexagon of the graphene sheet in the bilayer rGO, while the adsorbed water molecules are anchored to the epoxy groups on the monolayer rGO sheet, weakly binding with the epoxy oxygen atoms through the hydrogen bonding interaction. In this work, we consider the series of rGO films as gradually increasing the water content by controlling the number of water molecules, such as monolayer \ce{C72O12\cdot}6\ce{(H2O)} (water content = 9.3 wt\%) and bilayer \ce{C72O12$\cdot12$(H2O)} (17.0\%), \ce{C32O8\cdot}8\ce{(H2O)} (21.9\%), \ce{C32O12\cdot}12\ce{(H2O)} (27.3\%), \ce{C32O16\cdot}16\ce{(H2O)} (31.0\%). It is worth noting that the water contents of 21 and 31\% correspond to about 30 and 90\% RH respectively in the GO paper from the previous experimental works~\cite{Hatakeyama3,Paneri1,Paneri2}, indicating that the low, intermediate, and high humidity conditions are considered in this work.

We perform the variable cell relaxations of these water-intercalated rGO supercells allowing atoms to be relaxed while only atomic relaxations for the monolayer rGO supercells. The optimized atomic structures of the bilayer water-intercalated rGO sheets are shown in Figs.~\ref{fig_mostr}(a)$-$(d), where their interlayer distances and hydrogen bond lengths are also indicated. It is found that for the bilayer models the interlayer distance increases from 6.1 \AA~to 8.6 \AA~as a linear function of oxidation degree, being agreed well with those of the rGO films prepared via photoreduction process in experiment~\cite{Hatakeyama14}, as shown in Fig.~\ref{fig_mostr}(e). On the other hand, we find in the inset of Fig.~\ref{fig_mostr}(e) that the interlayer distance increases but the water binding energy decreases from 1.10 eV to 1.05 eV as both linear functions of water content. The increase of interlayer distance as increasing the water content is obvious given that the epoxy oxygen atoms attract water molecules through the hydrogen bonding interaction and thus more oxygen atoms can bind more water molecules, resulting in the expansion of interlayer space. In the case of monolayer \ce{C72O12\cdot}6\ce{(H2O)} model, the water binding energy is calculated to be 0.96 eV, being lower than those in the bilayer models.

The hydrogen bonds are observed between the water molecule themselves, which are of zigzag-type on the plane parallel to the basal graphene sheet as shown in Fig.~\ref{fig_mostr}(a), and between the epoxy oxygen atoms and water molecules. It is found that, while the hydrogen bond lengths between the water molecules are more or less invariable at the mean value of 1.9 \AA, those with the epoxy oxygen atoms increase gradually from 1.9 \AA~at the water content of 17.0\% to 2.0, 2.1, 2.4 \AA~at the water contents of 21.9, 27.3, 31.0\%. This gradual increasing tendency of hydrogen bond length between the epoxy oxygen atom and water molecule as increasing the water content is consistent with that of water binding energy, indicating a weakening of hydrogen bonding interaction at high humidity condition. It should be noted that those in the monolayer \ce{C72O12\cdot}6\ce{(H2O)} sheet are 1.7 \AA~for the former and 2.3 \AA~for the latter hydrogen bonds.

Figure~\ref{fig_dens} shows the electronic charge density difference when forming the water-adsorbed or water-intercalated rGO sheets. It is clear that the electronic charge transfer occurs upon the uptake of water into rGO sheets, where carbon and hydrogen atoms donate electrons while oxygen atoms receive them. This charge transfer becomes weakening as increasing the water content.

\subsection{In-plane proton transport}
\begin{figure}[!t]
\scriptsize
\begin{center}
(a)\includegraphics[clip=true,scale=0.22]{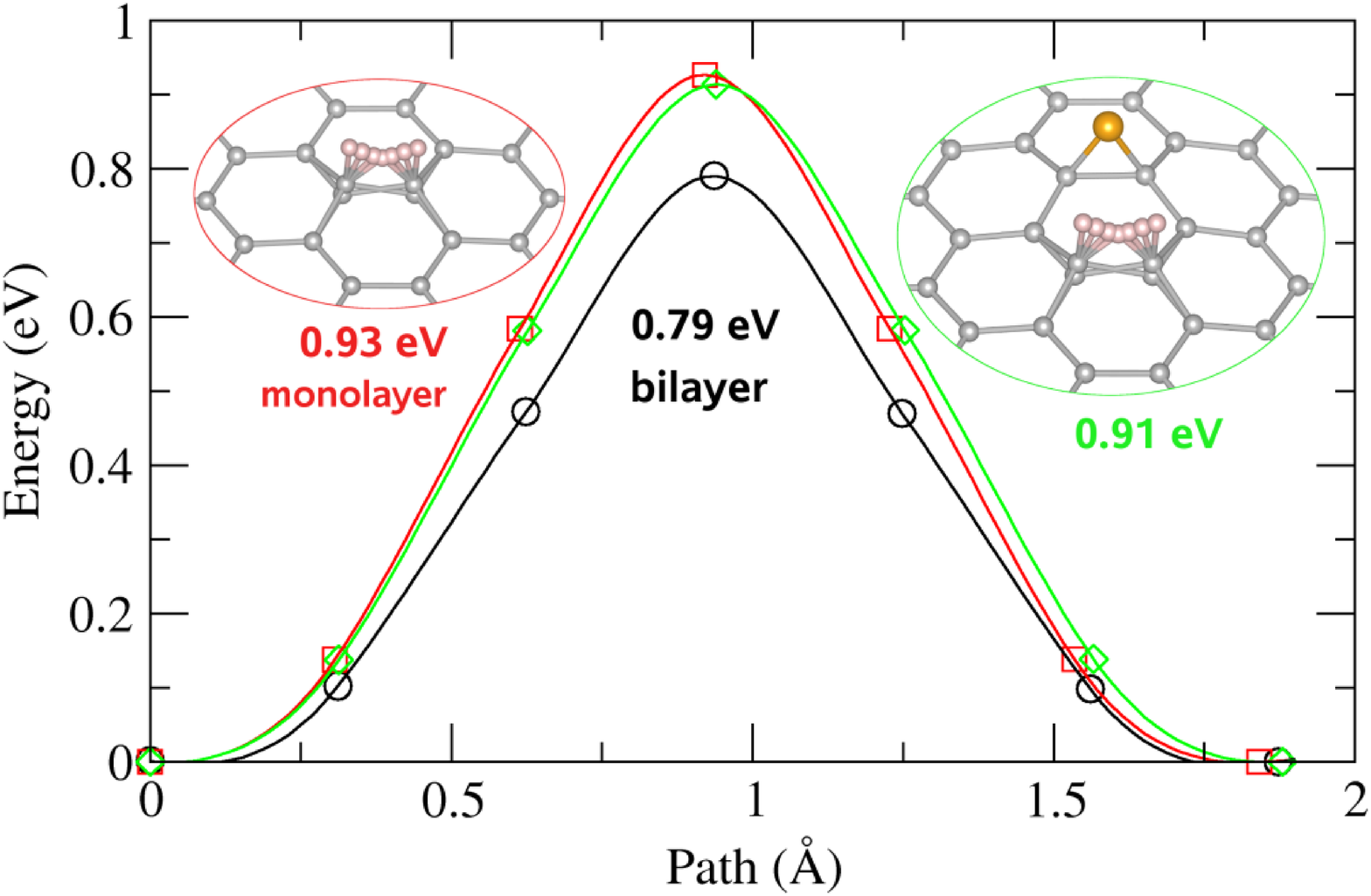} \\
(b)\includegraphics[clip=true,scale=0.22]{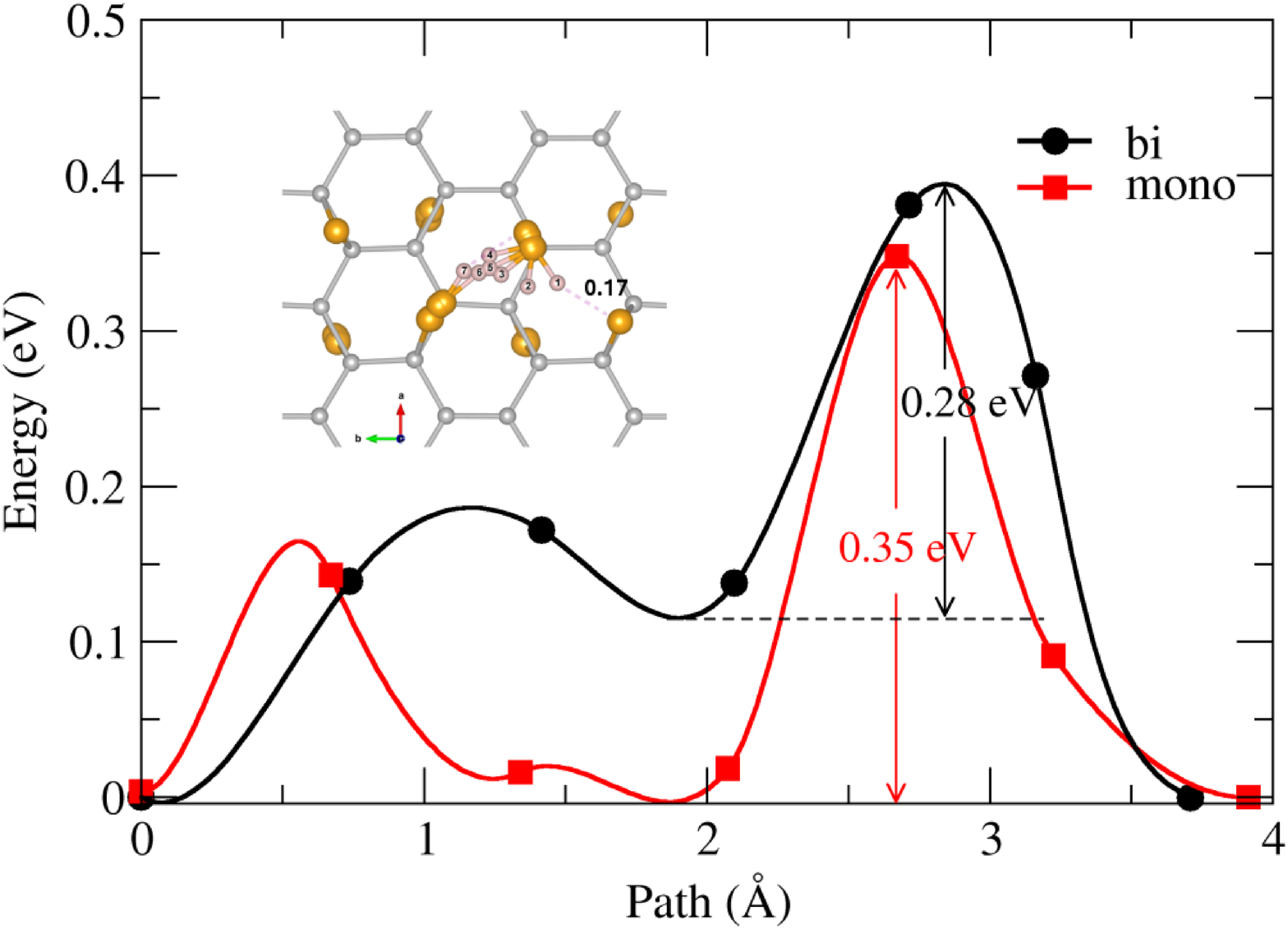}
\end{center}
\caption{\label{fig_nebg}Activation energy for proton migrations (a) along the C$-$C bond in monolayer (red), bilayer (black) graphene and monolayer graphene with one epoxy group (green), and (b) along the one-dimensional hydrogen bonded channel formed by epoxy and hydroxy groups in monolayer (red) and bilayer (black) \ce{C72O12} models. Inset shows the migration path with the hydrogen bond length (nm unit) in the bilayer rGO sheet.}
\end{figure}

Using the determined structures of models, we proceed to the in-plane proton transport on the rGO sheets. As a preliminary check for difficulty of the in-plane proton transport on the graphene sheet, we first consider the proton migration on the monolayer and bilayer graphene sheets. In these cases the proton can be adsorbed on the top of carbon atom with adsorption energies of $-$3.10 eV (monolayer) and $-$13.87 eV (bilayer), and migrates along the C$-$C bond with activation energies of 0.93 eV (monolayer), which is in good agreement with the previous result~\cite{Zhao}, and 0.79 eV (bilayer), as can be seen in Fig.~\ref{fig_nebg}(a). The existence of an epoxy group around the migration path slightly reduces the activation energy by 0.91 eV, possibly due to the hydrogen bonding interaction between the proton and oxygen atom.

We then considered the proton transport on the rGO sheet using \ce{C72O12} model, where the proton is adsorbed on the top of epoxy oxygen atom to form a hydroxy group~\cite{Yan09} with the adsorption energies of $-$7.37 and $-$14.47 eV for the monolayer and bilayer sheets. The proton of the hydroxy group is enforced to hop to the neighbouring epoxy oxygen atom ({\it i.e.} $-$OH $\rightarrow$ $-$O$-$) with the activation energies of 0.35 eV for the monolayer and 0.28 eV for the bilayer rGO sheets, as shown in Fig.~\ref{fig_nebg}(b). This confirms that the one-dimensional hydrogen-bonded channels formed by epoxy groups on the rGO sheet can remarkably reduce the activation energy for the in-plane proton transport~\cite{Wang,Topsakal,Dimiev,Raidongia}. In both cases of graphene and rGO sheets, bilayer sheets have lower activation energy than monolayer sheet, indicating that enhanced hydrogen-bonding interaction makes the proton transport fast.

\begin{figure}[!t]
\scriptsize
\begin{center}
(a)\includegraphics[clip=true,scale=0.23]{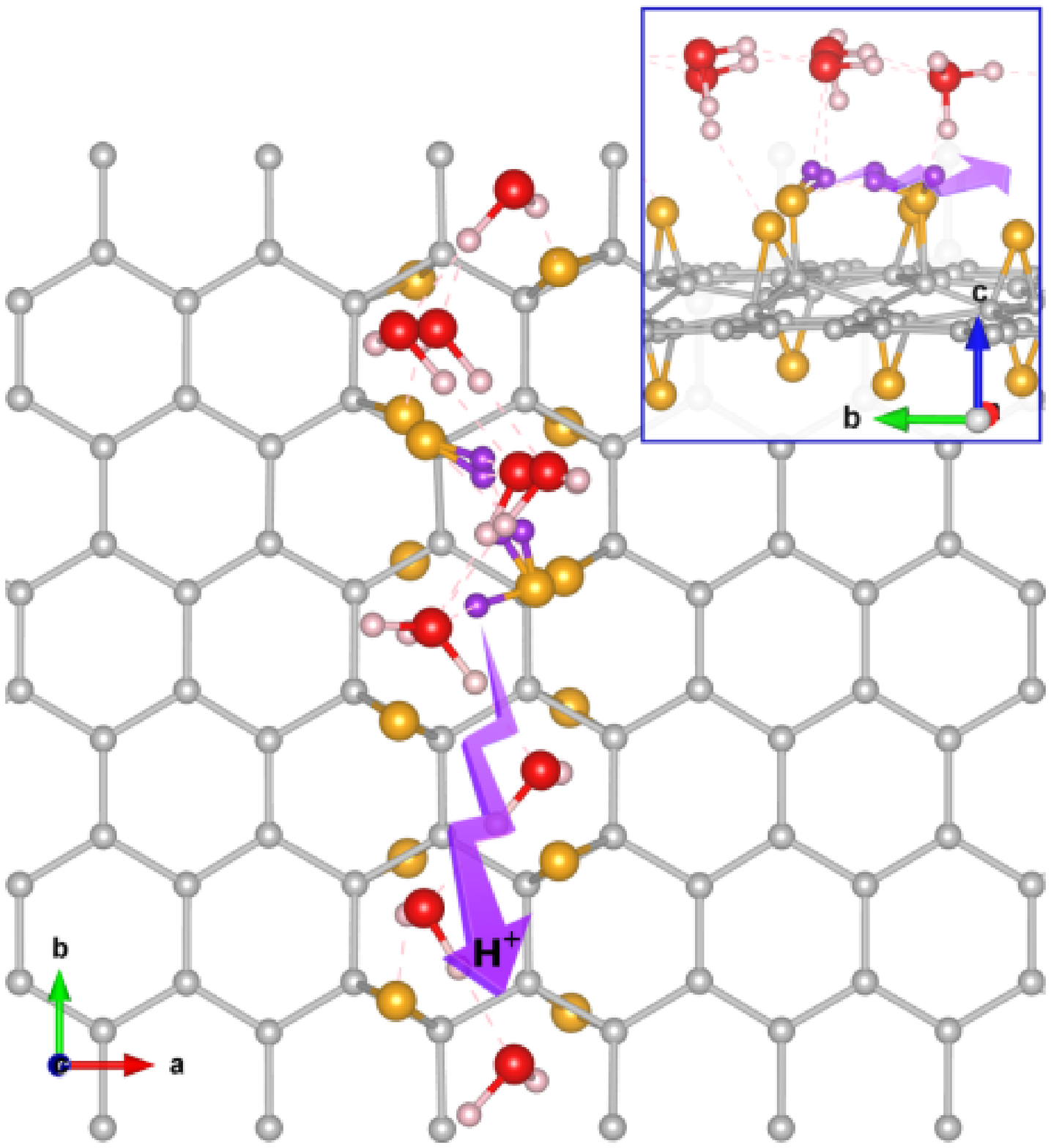}
(b)\includegraphics[clip=true,scale=0.23]{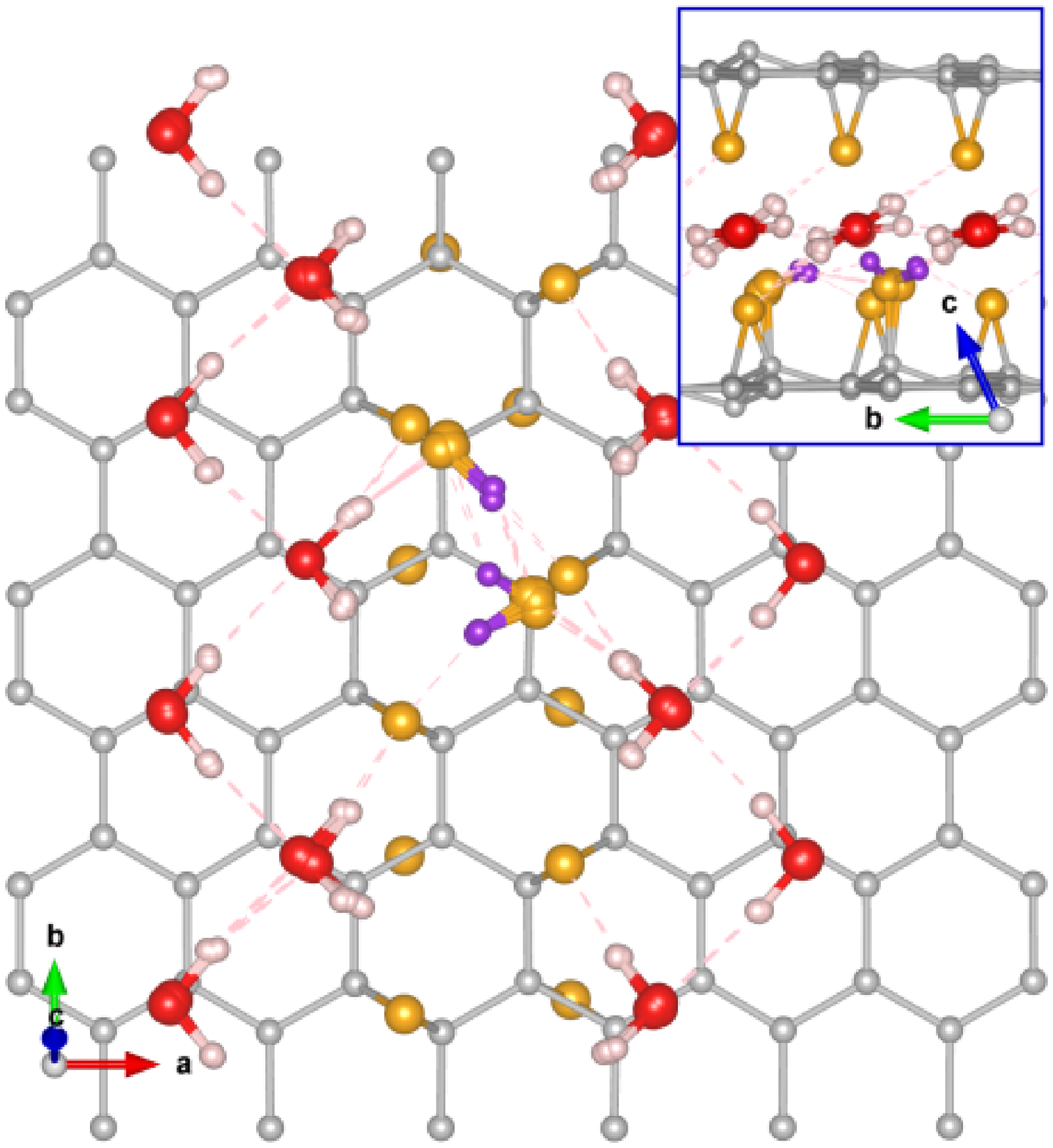} \\
(c)\includegraphics[clip=true,scale=0.23]{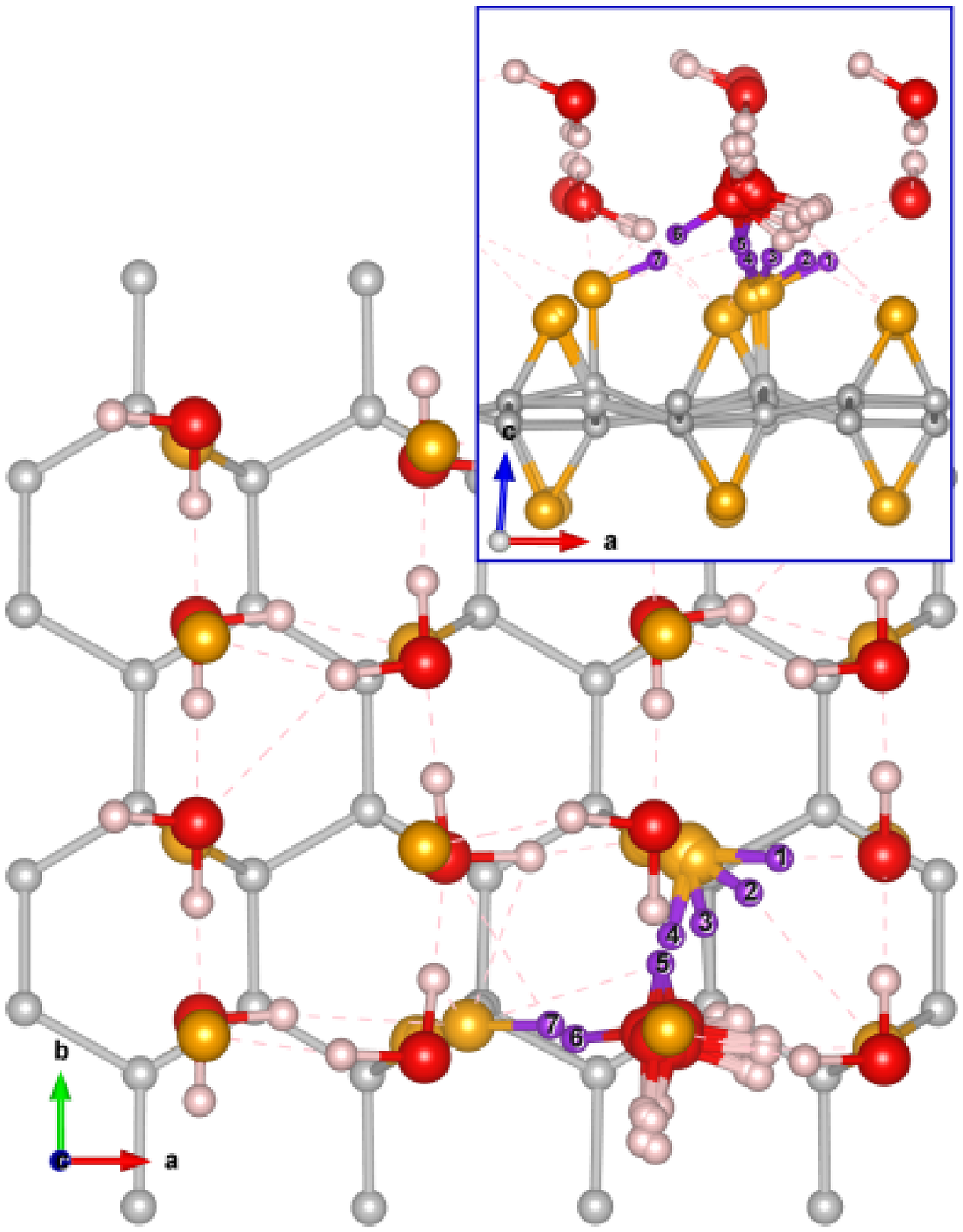}
(d)\includegraphics[clip=true,scale=0.27]{fig5d.eps}
\end{center}
\caption{\label{fig_migepo}Top view of migration paths for epoxy-mediated proton transport in (a) monolayer \ce{C72O12\cdot}6\ce{(H2O)}, (b) bilayer \ce{C72O12\cdot}12\ce{(H2O)} and (c) bilayer \ce{C32O16\cdot}16\ce{(H2O)} rGO sheets. Insets show the perspective view, and arrows indicate the path. (d) Activation energies for these proton migrations.}
\end{figure}

For the cases of water-adsorbed or water-intercalated rGO sheets, it is not easy to clarify the adsorption sites and migration paths of proton. Here, we propose two different transport mechanisms, namely, epoxy-mediated and water-mediated proton hoppings, in which protons on the hydroxy group ($-$OH) for the former mechanism or hydronium group ($-$\ce{OH3}) for the latter are hopping from one group site to another. It should be noted that the concept of proton transport by hopping is similar to the case in water via Grotthuss mechanism and in Nafion through the sulfonic acid (\ce{SO3H})~\cite{HuPhd}. In addition, we regard that a mixing mechanism, {\it i.e.} epoxy-water-mediated proton transport, is not ruled out. In the former case, we follow the same way as in the case of above mentioned anhydrous rGO sheet, whereas in the latter case we perform $\Delta$BVS analysis to predict the adsorption sites and migration paths of proton.

\begin{figure}[!t]
\scriptsize
\begin{center}
(a)\includegraphics[clip=true,scale=0.29]{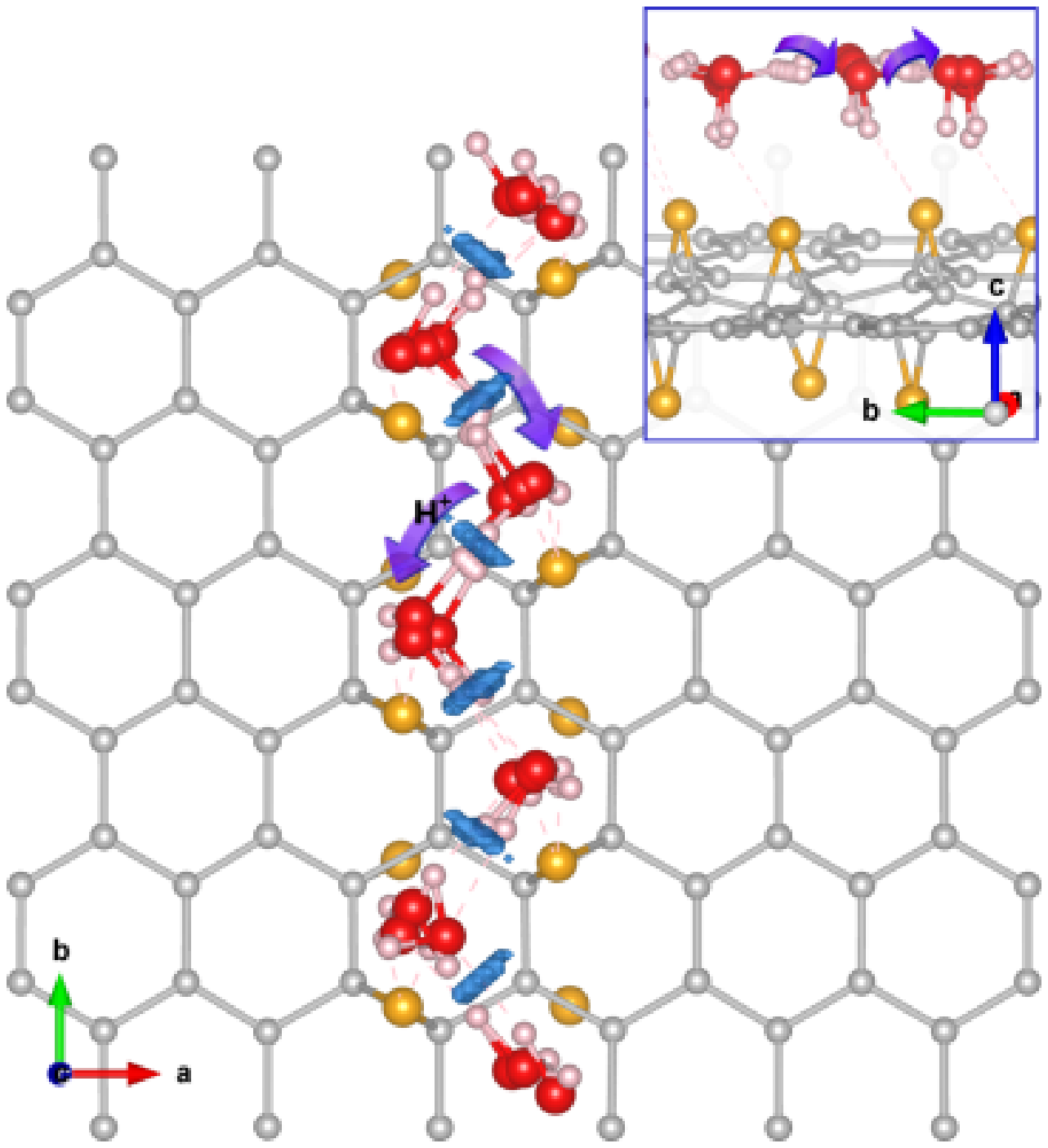}
(b)\includegraphics[clip=true,scale=0.29]{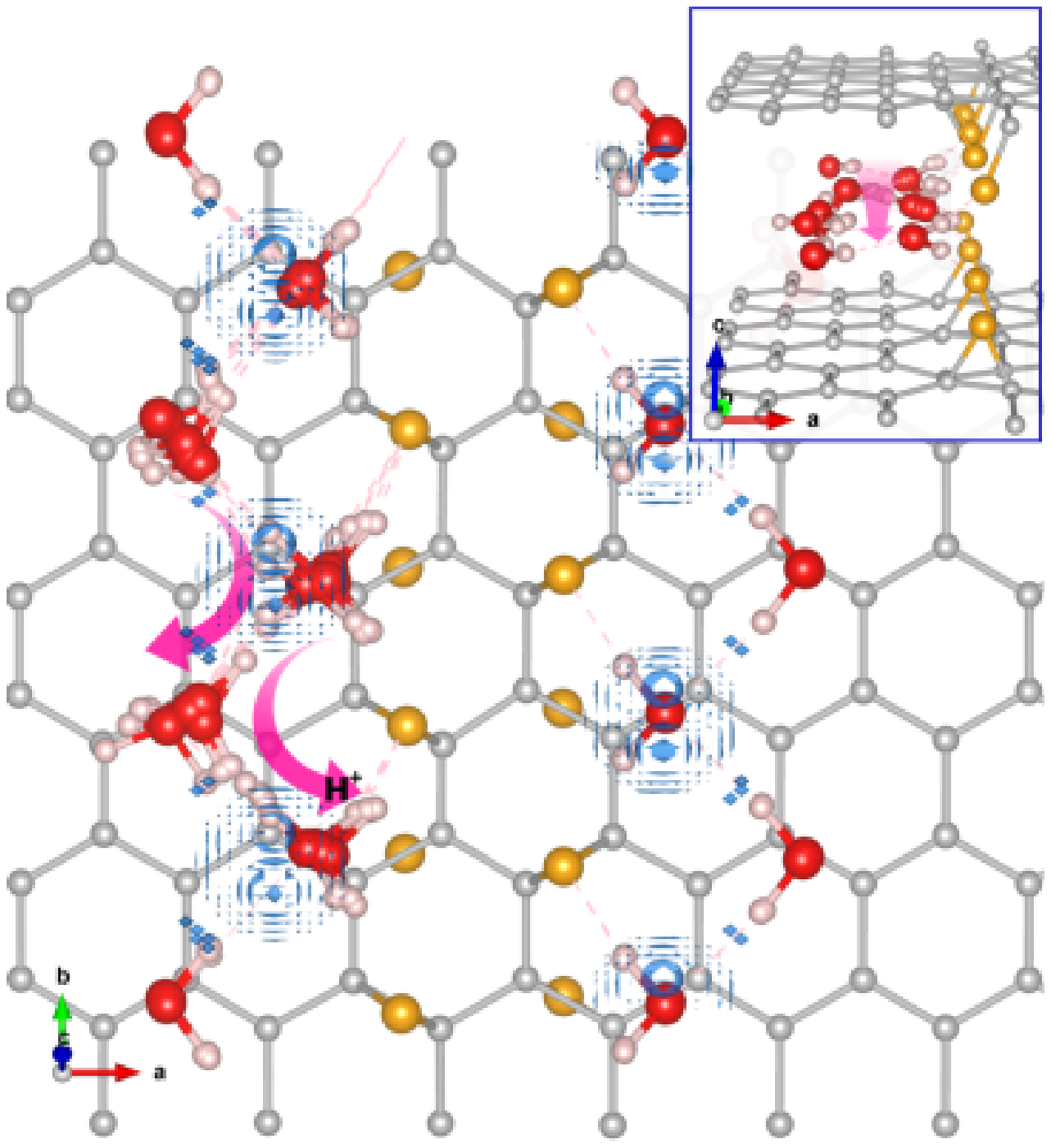} \\
(c)\includegraphics[clip=true,scale=0.27]{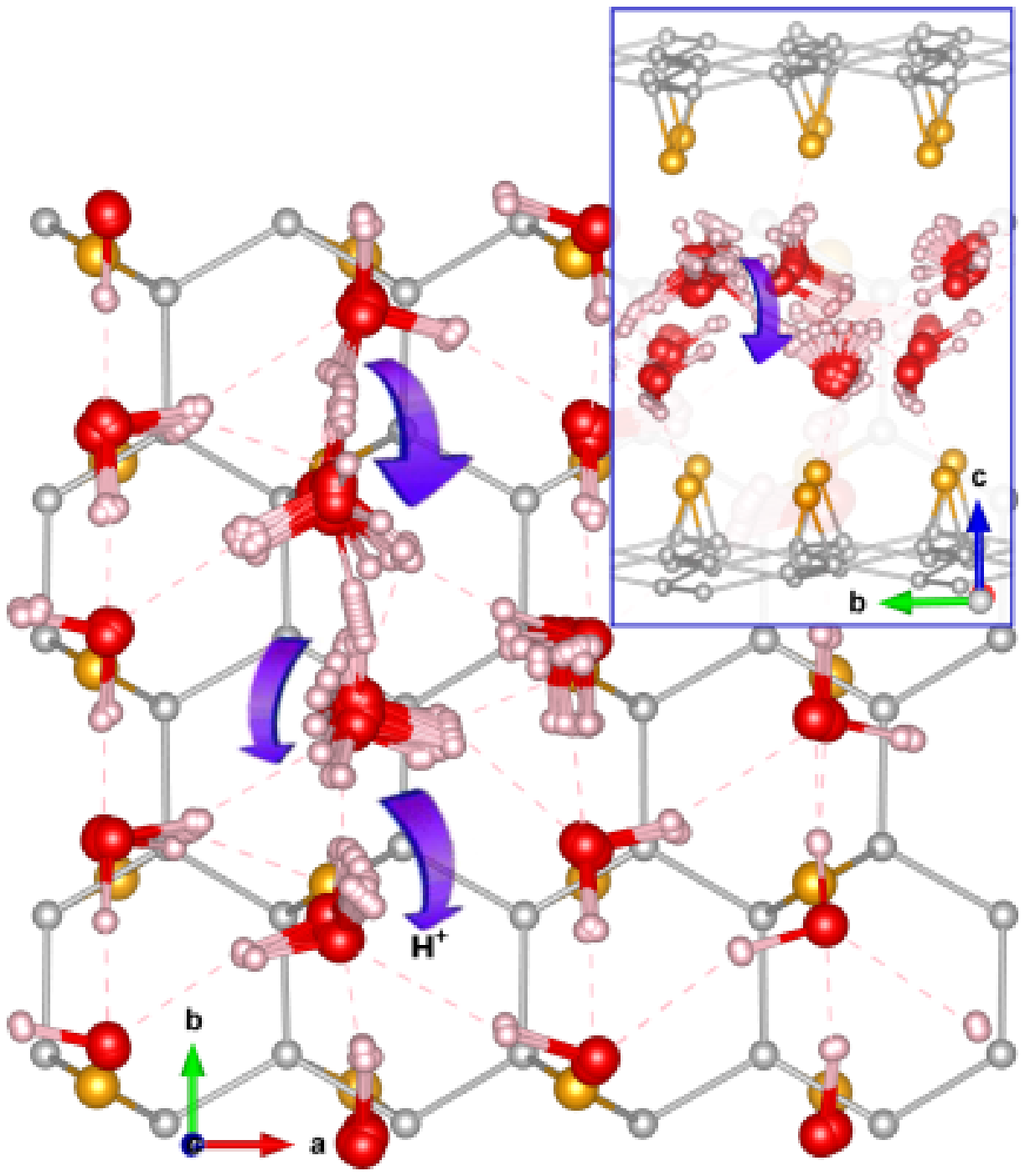}
(d)\includegraphics[clip=true,scale=0.37]{fig6d.eps}
\end{center}
\caption{\label{fig_migwat}Top view of migration paths for water-mediated proton transport in (a) monolayer \ce{C72O12\cdot}6\ce{(H2O)}, (b) bilayer \ce{C72O12\cdot}12\ce{(H2O)} and (c) \ce{C32O16\cdot}16\ce{(H2O)}. Insets show the perspective view, and blue isosurfaces in (a) and (b) represent the $\Delta$BVS at the value of 3. (d) Activation energies for these proton migrations.}
\end{figure}

The activation barrier for the epoxy-mediated proton migration in the monolayer \ce{C72O12\cdot}6\ce{(H2O)} model is determined to be 0.21 eV, which is lower than that in its anhydrous counterpart \ce{C72O12} model (0.35 eV) (Fig.~\ref{fig_migepo}). Such enhancement of proton migration is attributed to hydrogen bonding interaction between the proton and adsorbed water molecules that are placed over epoxy groups, forming a hydrogen-bonded water channel. On the contrary, the activation energy in the case of bilayer \ce{C72O12\cdot}12\ce{(H2O)} model is determined to be higher as 0.63 eV. By inspecting the migration path, we find out that unlike the former case there is no hydrogen bond between the proton and intercalated water molecule in this model, in which water molecules are placed interlayer space away from the row of epoxy groups. Meanwhile, in the case of bilayer \ce{C32O16\cdot}16\ce{(H2O)} the proton migration is realized according to the mixed way of epoxy-water-mediated mechanism, although we enforce the epoxy-mediated migration. Here, the activation energies are estimated to be 0.48 eV for proton movement from water to epoxy group, which occurs without hydrogen bonding interaction, and 0.37 eV along the epoxy-mediated path with the effect of hydrogen bond. These results indicate that the hydrogen bonding interaction between the proton and water molecule can enhance the in-plane proton transport.

Finally, we present the results for the water-mediated proton transports in the monolayer \ce{C72O12\cdot}6\ce{(H2O)} and bilayer \ce{C72O12\cdot}12\ce{(H2O)} and \ce{C32O16\cdot}16\ce{(H2O)} models in Figs.~\ref{fig_migwat}(a)$-$(c). As mentioned above, water molecules are connected with their neighbors by hydrogen bond, forming zigzag-type two-dimensional channel on the top of epoxy group in the case of monolayer \ce{C72O12\cdot}6\ce{(H2O)} and bilayer \ce{C32O16\cdot}16\ce{(H2O)} or in the interlayer space in the case of \ce{C72O12\cdot}12\ce{(H2O)}. As clarified by $\Delta$BVS analysis, the inserted proton attaches to the water molecule with the adsorption energies of $-$7.39, $-$13.11 and $-$12.76 eV in these sheets, forming hydronium ion (\ce{H3O+}), and then the nearest one of its three hydrogen atoms moves to the neighbouring water molecule. Rotation of water molecules to some degree is observed during the proton hopping. As shown in Fig.~\ref{fig_migwat}(d), the corresponding activation energies are calculated to be 0.16, 0.17 and 0.23 eV in these hydrous rGO sheets, being much lower than along the epoxy-mediated paths and in the anhydrous GO sheet. Moreover, these are comparable with the experimental value of 0.12 eV~\cite{Hatakeyama1}. The similar values in the monolayer \ce{C72O12\cdot}6\ce{(H2O)} and the bilayer \ce{C72O12\cdot}12\ce{(H2O)} can be explained by the similar water-mediated paths, giving an evidence of indirect effect of epoxy groups, which play a role of holding the water molecules. Meanwhile, slightly higher value in the bilayer \ce{C32O16\cdot}16\ce{(H2O)} indicates that too many water molecules around the path may disturb the proton hopping due to an attraction of the proton by another water molecule through hydrogen bonding interaction.

\begin{table}[!t]
\caption{\label{table_str}Overview of calculation data for rGO and the series of water-adsorbed rGO models with oxidation degree (O/(C+O)) and water content in weight percentage. Given are the interlayer distance ($d$) and activation energy ($E_\text{a}$).}
\small
\begin{ruledtabular}
\begin{tabular}{lccccc}
      & Layer & O/(C+O) &  Water  & $d$    & $E_\text{a}$\\
Model & type  & (\%)    & (wt.\%) & (nm) & (eV) \\
\hline 
\ce{C72O12}                  & mono & 14.3 & $-$  & $-$  & 0.35 \\
\ce{C72O12\cdot}6\ce{(H2O)}  & mono & 14.3 & 9.3  & $-$  & 0.16 \\
\ce{C72O12\cdot}12\ce{(H2O)} & bi   & 14.3 & 17.0 & 0.61 & 0.17 \\
\ce{C32O8\cdot}8\ce{(H2O)}   & bi   & 20.0 & 21.9 & 0.71 & -    \\
\ce{C32O12\cdot}12\ce{(H2O)} & bi   & 27.3 & 27.3 & 0.76 & -    \\
\ce{C32O16\cdot}16\ce{(H2O)} & bi   & 33.3 & 31.0 & 0.86 & 0.23 \\
\end{tabular}
\end{ruledtabular}
\end{table}

In Table~\ref{table_str}, we summarize the main result for rGO and hydrous rGO models with their oxidation degrees and water contents, including the interlayer distance in the bilayer models and the activation energy for the in-plane proton transport. It is revealed that the in-plane proton conductivity is enhanced through the hydrogen bonding interaction between the proton and water when mediated by water at humid condition. There are several options to further enhance the in-plane proton conductivity, {\it e.g}, by functionalizing GO films with sulfonic acid group. This work may contribute to the development of efficient solid proton exchange membranes based on GO films.

\section{Conclusions}
In conclusion, we have studied the atomic structures of water-adsorbed monolayer and water-intercalated bilayer rGO sheets, varying the oxidation degree and water content, and calculated the activation energies for the in-plane proton transports using the first-principles method. Our calculations have been shown to offer good agreement with the experimental measures for the interlayer distances of bilayer series as increasing the oxidation degree, shedding light on the hydrogen bond between water molecule and epoxy group. We suggest that in these hydrous rGO sheets the proton can hop along two different mechanisms, epoxy-mediated and water-mediated paths, and conclude that the water-mediated proton transport is more likely to occur due to its much lower activation energy (0.16, 0.17 eV) compared with the epoxy-mediated transports and close value to the experiment. Our study may contribute to the understanding of the proton conductivity enhancement of rGO films at humid condition, and reveals new prospects for developing efficient solid proton exchange membranes based on GO films.

\section*{Acknowledgments}
This work is supported as part of the fundamental research project ``Design of Innovative Functional Materials for Energy and Environmental Application'' (no. 2016-20) funded by the State Committee of Science and Technology, DPR Korea. Computation was done on the HP Blade System C7000 (HP BL460c) that is owned by Faculty of Materials Science, Kim Il Sung University.

\bibliographystyle{apsrev}
\bibliography{Reference}

\end{document}